\documentclass[papera4,12pt]{elsarticle}

\usepackage{setspace}
\usepackage{multirow}
\usepackage{graphicx}
\usepackage{longtable}
\usepackage{threeparttable}
\usepackage{amsmath}
\usepackage{amssymb}
\usepackage{lineno}
\usepackage{color}
\usepackage{amsmath}
\usepackage{braket}
\usepackage{natbib}
\setcitestyle{numbers,sort&compress}

\begin{document}
\begin{frontmatter}
\title{Thermodynamical and topological properties of metastable Fe$_3$Sn}

\author[a]{Ilias Samathrakis\corref{cor1}}
\author[a]{Chen Shen\corref{cor1}}
\author[a,b]{Kun Hu\fnref{label2}}
\author[a]{Harish K. Singh}
\author[a]{Nuno Fortunato}
\author[b]{Huashan Liu}
\author[a]{Oliver Gutfleisch}
\author[a]{Hongbin Zhang\fnref{label2}}

\affiliation[a]{School of Materials Science, Technical University Darmstadt,Darmstadt,64287,Germany}
            
\affiliation[b]{School of Materials Science and Engineering, Central South University,Changsha,410083,China}

\cortext[cor1]{These authors contributed equally to this work.}
\fntext[label2]{Corresponding authors: kh89jyni@tu-darmstadt.de (Kun Hu), hzhang@tmm.tu-darmstadt.de (Hongbin Zhang)}

\begin{abstract}
Combining experimental data, first-principles calculations, and Calphad assessment, thermodynamic and topological transport properties of the Fe-Sn system were investigated. Density functional theory (DFT) calculations were performed to evaluate the intermetallics' finite-temperature heat capacity (C$_p$). A consistent thermodynamic assessment of the Fe-Sn phase diagram was achieved by using the experimental and DFT results, together with all available data from previous publications. Hence, the metastable phase Fe$_3$Sn was firstly introduced into the current metastable phase diagram, and corrected phase locations of Fe$_5$Sn$_3$ and Fe$_3$Sn$_2$ under the newly measured corrected temperature ranges. Furthermore, the anomalous Hall conductivity and anomalous Nernst conductivity of Fe$_3$Sn were calculated, with magnetization directions and doping considered as perturbations to tune such transport properties. It was observed that the enhanced anomalous Hall and Nernst conductivities originate from the combination of nodal lines and small gap areas that can be tuned by doping Mn at Fe sites and varying magnetization direction. 
\end{abstract}



\end{frontmatter}
\renewcommand{\baselinestretch}{2}
\section{\label{INTRO}Introduction}
The kagome lattice is a 2D network of corner-sharing triangles that has been intensively investigated the last years. Due to its unusual geometry, it offers a playground to study interesting physics including frustrated, correlated~\cite{yin2018giant,fenner2009non}, exotic topological quantum~\cite{
mekata2003kagome,zhou2017quantum,ohgushi2000spin,yan2011spin,
han2012fractionalized,
mazin2014theoretical,
chisnell2015topological,
xu2015intrinsic,
zhu2016interaction,
yan2011spin,
yin2018giant,
guo2009topological,
mazin2014theoretical,
xu2011chern,
Chen:2014,
Kubler:2014,
tang2011high}, topological Chern~\cite{kane2005quantum}, insulating and Weyl semimetal~\cite{xu2011chern} phases, originating from the interplay between magnetism and electronic topology. In fact, the kagome lattice has been realized in several materials including metal stannides, germannides~\cite{giefers2006high,haggstrom1975investigation} as well as T$_m$X$_n$ compounds with T=Mn, Fe, Co, X=Sn, Ge ($m$:$n$=3:1, 3:2, 1:1)~\cite{kang2020dirac}. Recent studies demonstrated that Fe-Sn-based kagome compounds exhibiting interesting properties, such as large magnetic tunability~\cite{yin2018giant}. Furthermore they can host Dirac fermions and flat bands, as found in Fe$_3$Sn$_2$~\cite{lin2018flatbands,lin2020tunable} and FeSn~\cite{lin2020dirac,kang2020dirac}. The existence of spin degenerate band touching points was linked to the generation of several interesting phenomena. Specifically, the anomalous Hall effect (AHE) results in a transverse spin polarized charge current (charge current and spin current due to the imbalance of spin up and spin down electrons in ferromagnets) in response to a longitudinal charge current, in the absence of an external magnetic field~\cite{wang2016anomalous,tanaka2020three,li2019large,kida2011giant,li2020large}. This applies also to its thermal counterpart, the anomalous Nernst effect (ANE), in which the external stimuli is replaced by a thermal gradient~\cite{zhang2021topological} as well as the Seebeck effect~\cite{sales2019electronic}.

Interestingly, the Fe-Sn-based intermetallics compounds not only exhibit attractive topological transport properties, but also show rich magnetic properties. 
In our previous studies~\cite{fayyazi2017bulk,fayyazi2019experimental}, a DFT screening of the Fe-Sn phase diagram was used to identify Fe-Sn based phases with potential to be stabilized upon alloying, and their magnetization and magnetocrystalline anisotropy were evaluated. 
The results revealed that a strong anisotropy as observed in Fe$_3$Sn may also be found in other Fe-Sn based phases, having high potential to be used as hard magnetic materials.
Meanwhile, we applied the reactive crucible melting (RCM) approach to the Fe-Sn binary system, and observed 3 metastable intermetallic compounds, namely Fe$_3$Sn, Fe$_5$Sn$_3$, Fe$_3$Sn$_2$, which are ferromagnetic and exist between 873 K and 1173 K. We found that such metastable phases can be synthesized using the RCM method at specific temperature ranges. Furthermore, we speculted that the phase diagram reported in the literature~\cite{giefers2006high,treheux1974etude,kumar1996} is inaccurate in the temperature interval 750-765 $^\circ$C as Fe$_3$Sn can exist at 750 $^\circ$C.
Therefore, to further explore the interesting properties of metastable Fe-Sn phases, it is important to understand the phase diagram and thermodynamical properties of the Fe-Sn system.

In this work, we adopted our new measurements~\cite{fayyazi2017bulk,fayyazi2019experimental} on the equilibria states of Fe$_3$Sn, Fe$_5$Sn$_3$, Fe$_3$Sn$_2$, combined with the thermodynamic properties of such intermetallic phases obtained based on first-principles calculations. A consistent thermodynamic assessment of the Fe-Sn system was then developed based on all available experimental and first-principles results. Furthermore, the AHC and ANC of Fe$_3$Sn were calculated and its dependence on the magnetization direction and doping were evaluated. 
We observed that there exist significant changes in AHC and ANC by tuning the Fermi energy via Mn-doping. Therefore, Fe$_3$Sn renders itself as a promising candidate for novel transverse thermoelectric devices with potential applications. Additionally, we suspect high-throughput calculations~\cite{zhang2021high,shen2021designing} can be performed to search for more intriguing magnetic intermetallic compounds with singular topological transport properties, assisted by automated Wannier function construction~\cite{Zeying:2018} for transport property calculations.

\section{Methodology}

\subsection{First-principles calculations}
Our calculations were performed using the generalized gradient approximation (GGA) for the exchange-correlation functional, in the parameterization of Perdew-Burke-Ernzerhof~\cite{Perdew:1996} for the Vienna ab$-$initio Simulation Package (VASP) ~\cite{kresse1996efficiency,togo2015first}.
The energy cutoff is set at 600 eV and at least 5000 k-points in the first Brillouin zone with $\Gamma$-centered k-mesh were used for the hexagonal lattices(Fe$_3$Sn, FeSn, and Fe$_5$Sn$_3$), while for all the other structures, Monkhorst-Pack grids were used. The energy convergence criterion was set as $10^{-6}$ eV, and $10^{-5}$ eV$/\AA$ was set as the tolerance of forces during the structure relaxation. The enthalpy of formation, $\Delta_fH(Fe_xSn_y)$, for the Fe$_x$Sn$_y$ intermetallic compounds was obtained following

\begin{equation}
\label{eq:hf}
\Delta_fH(Fe_xSn_y)=E_{Fe_xSn_y}-\frac{x}{x+y}E_{Fe}-\frac{y}{x+y}E_{Sn},
\end{equation}
where all the total energies for the equilibrium phases in their corresponding stable structures were obtained after structural relaxation. 

For the phonon calculations, the frozen phonon approach was applied using the PHONOPY package~\cite{togo2015first}. The temperature-dependent thermodynamical properties were calculated by using the quasi-harmonic approximation~\cite{baroni2010density}.
The Gibbs free energy $G(T,P)$ at temperature $T$ and pressure $P$ can be obtained from the Helmholtz free energy $F(T,V)$ as follows: \cite{hu2017first}
    \begin{equation}
        G(P,T)-PV =F(T,V) =E_{0}(V)+F_{vib}(V,T)+F_{el}(V,T)+ F_{magn}(V,T),
        \label{eq:GPT}
    \end{equation}
where $E_{0}$(V) is the total energy at zero Kelvin without the zero-point energy contribution, which were determined by fitting of the energies with respect to the volume data using the Brich-Murnaghan equation of state (EOS)~\cite{wang2004thermodynamic}. 
$F_{vib}$ corresponds to the lattice vibration contribution to the Helmholtz energy, which can be derived from the phonon density of states (PhDOS), $g(\omega,V)$, by using the following equation~\cite{hu2017first}:
    \begin{equation}
         F_{vib}(V,T)=k_{B}T\int_{0}^{\infty}\ln\ [ 2 \sinh\frac{\hbar\omega}{2k_{B}T}]g(\omega,V)d\omega,
    \end{equation}
where $k_{B}$ and $\hbar$ are the Boltzmann constant and reduced Planck constant, respectively, and $\omega$ denotes the phonon frequency for a given wave vector $q$. The PhDOS $g(\omega,V)$ can be obtained by integrating the phonon dispersion in the Brillouin zone.
The third term $F_{el}$ represents the electronic contribution to the Helmholtz free energy, obtained by~\cite{liang2019phase}:
\begin{equation}
F_{el}(V,T)=E_{el}(V,T)-T\cdot S_{el}(V,T)
\end{equation}
where $E_{el}(V,T)$ and $S_{el}(V,T)$ indicate the electronic energy and electronic entropy, respectively. With the electronic $DOS$, both terms can be formulated as~\cite{liang2019phase}:
    \begin{equation}
        E_{el}(V,T)=\int n\left (  \epsilon \right )f\epsilon d\epsilon -\int_{-\infty }^{\epsilon_{F}}n(\epsilon,V)d\epsilon,
    \end{equation}
    \begin{equation}
        S_{el}(V,T)=-k_{B}\int n\epsilon [ flnf +(1-f)ln(1-f)]d\epsilon, 
    \end{equation}
where $n(\epsilon)$ is the electronic $DOS$, $f$ represents the Fermi-Dirac distribution function and $\epsilon_{F}$ is the Fermi energy.

Finally, based on the Inden-model~\cite{inden1976approximate}, the magnetic Gibbs energy can be formulated as:
    \begin{equation}
        F_{magn}=RTln(\beta ^{\varphi }+1)f(\tau ), \tau =T/T_{c}^{\varphi},
    \end{equation}
where $T_{c}^{\varphi}$ is the Curie temperature for the phase-$\varphi$. $\beta^{\varphi }$ is the average magnetic moment per atom.

In order to evaluate AHC, we projected the Bloch wave functions onto maximally localized wannier functions (MLWF) using Wannier90, following Ref.~\cite{Mostofi:2008}. A total number of 124 MLWFs, originating from the \textit{s}, \textit{p} and \textit{d} orbitals of Fe atoms and the s and p orbitals of Sn atoms, are used. AHC is obtained by integrating the Berry curvature according to the formula:
\begin{equation}
\begin{split}
& \sigma_{\alpha \beta} = -\frac{e^2}{\hbar} \int \frac{d\mathbf{k}}{\left( 2\pi \right)^3} \sum f \left[ \epsilon\left(\mathbf{k}\right)-\mu\right] \Omega_{n,\alpha \beta} \left(\mathbf{k}\right), \\
& \Omega_{n,\alpha \beta} \left(\mathbf{k}\right) = -2 Im \sum_{m \neq n} \frac{\braket{\mathbf{k}n|v_{\alpha}|\mathbf{k}m}\braket{\mathbf{k}m|v_{\beta}|\mathbf{k}n}}{\left[\epsilon_m\left(\mathbf{k}\right)-\epsilon_n\left(\mathbf{k}\right)\right]^2},
\label{bc-ahc}
\end{split}
\end{equation}
with $\mu$, $f$, $n$, $m$, $\epsilon_n\left(\mathbf{k}\right)$, $\epsilon_m\left(\mathbf{k}\right)$ and $v_{\alpha}$  being the Fermi level, the Fermi-Dirac distribution function, the occupied Bloch band, the empty Bloch band, their corresponding energy eigenvalues and the Cartesian component of the velocity operator. The integration is performed on a $270\times270\times350$ mesh using Wanniertools~\cite{wanniertools:2018}. ANC is evaluated using an in-house developed Python script, following the formula:
\begin{equation}
a_{\alpha \beta} = -\frac{1}{e} \int d\epsilon \frac{\partial f}{\partial \mu} \sigma_{\alpha \beta} \left(\epsilon\right) \frac{\epsilon-\mu}{T}\\
\end{equation}
where T, e and $\epsilon$ are the temperature, the electronic charge and the energy point within the integration energy window respectively. An energy grid of 1000 points within the window $\left[-0.5,0.5\right]$ with respect to the Fermi level was chosen.

\subsection{CALPHAD modeling}
\subsubsection{Pure elements}
The Gibbs free energies for pure Fe and Sn were taken from the Scientific Group Thermodata Europe (SGTE) pure element database~\cite{dinsdale1991sgte}, which was described by:
\begin{equation}
\scriptsize
^{\circ}\textrm{G}_i^{\phi}(T)=G_i^{\phi}(T)-H_{i,SER}(298.15~K)=a+bT+cTln(T)+dT^2+eT^3+fT^{-1}+gT^7+hT^{-9}~,
\end{equation}
where \textit{i} represents the pure elements Fe or Sn, $H_{i,SER}(298.15~K)$ is the molar enthalpy of element i at 298.15 K in its standard element reference (SER) state, and \textit{a} to \textit{h} are known coefficients. 

\subsubsection{Solution Phases}
The solution phases, Liquid, BCC$\_$A2, FCC$\_$A1 and BCT$\_$A5 phases are described using the substitutional solution model, with the corresponding molar Gibbs free energy formulated as:
    \begin{equation}
        G_{m}^{\varphi}=x_{Fe}G_{Fe}^{\varphi }(T)+x_{Sn}G_{Sn}^{\varphi }(T)+RT(x_{Fe} lnx_{Fe}+x_{Sn}lnx_{Sn})+G^{ex}+G^{magn},      
    \end{equation}
where $x_{Fe}$ and $x_{Sn}$ are the mole fraction of Fe and Sn in the solution, respectively. Taken from SGTE~\cite{dinsdale1991sgte}, $G_{i}^{\varphi }$ denotes the molar Gibbs free energy of pure Fe and Y in the structure $\varphi$ at the given temperature. $G^{ex}$ denotes the excess Gibbs energy of mixing, which measures the deviation of the actual solution from the ideal solution behaviour, modelled using a Redlich-Kister polynomial \cite{redlich1948algebraic}:
    \begin{equation}
          G^{ex}=x_{Fe}x_{Sn}\sum_{j=0}^{n} {^{(j)}{L}_{Fe,Sn}^{\varphi}(x_{Fe}-x_{Sn})^{j}}. 
    \end{equation}
The ${\it j}-th$ interaction parameter between Fe and Sn is described by $_{}^{(j)}\textrm{L}_{Fe,Y}^{\varphi }$, which is modelled in terms of a$^*$+b$^*$T.

\subsubsection{Stoichiometric intermetallic compounds}
Fe$_5$Sn$_3$, Fe$_3$Sn$_2$, Fe$_3$Sn, FeSn, and FeSn$_2$ were considered as stoichiometric phases.
The Gibbs free energies per mole atom of these phases were thus expressed as follows:

\begin{equation}
G_m^{Fe_xSn_y}=\frac{x}{x+y}G_{Fe,SER}+\frac{y}{x+y}G_{Sn,SER}+\Delta G^{Fe_xSn_y}_f(T)~,
\end{equation}
where $\Delta G^{Fe_xSn_y}_f(T)$ is the Gibbs free energy of formation of the stoichiometric compound Fe$_x$Sn$_y$ which can be expressed as:

\begin{equation}
\Delta G^{Fe_xSn_y}_f(T)=A_3+B_3T~,
\label{a}
\end{equation}
where the coefficients \textit{A$_3$}, \textit{B$_3$} are the parameters to be optimized.
Since there is no experimental data of the thermodynamic properties for such intermetallic phases, the calculated enthalpies of formation for these phases from DFT calculations were treated as initial values of the coefficient \textit{A$_3$} in Eq.~\ref{a} in the present optimization.

\section{\label{RES}Results and Discussion}
\subsection{Metastable phase diagram}
\begin{figure}
    \centering   
    \includegraphics[width=\textwidth,height=0.9\textheight]{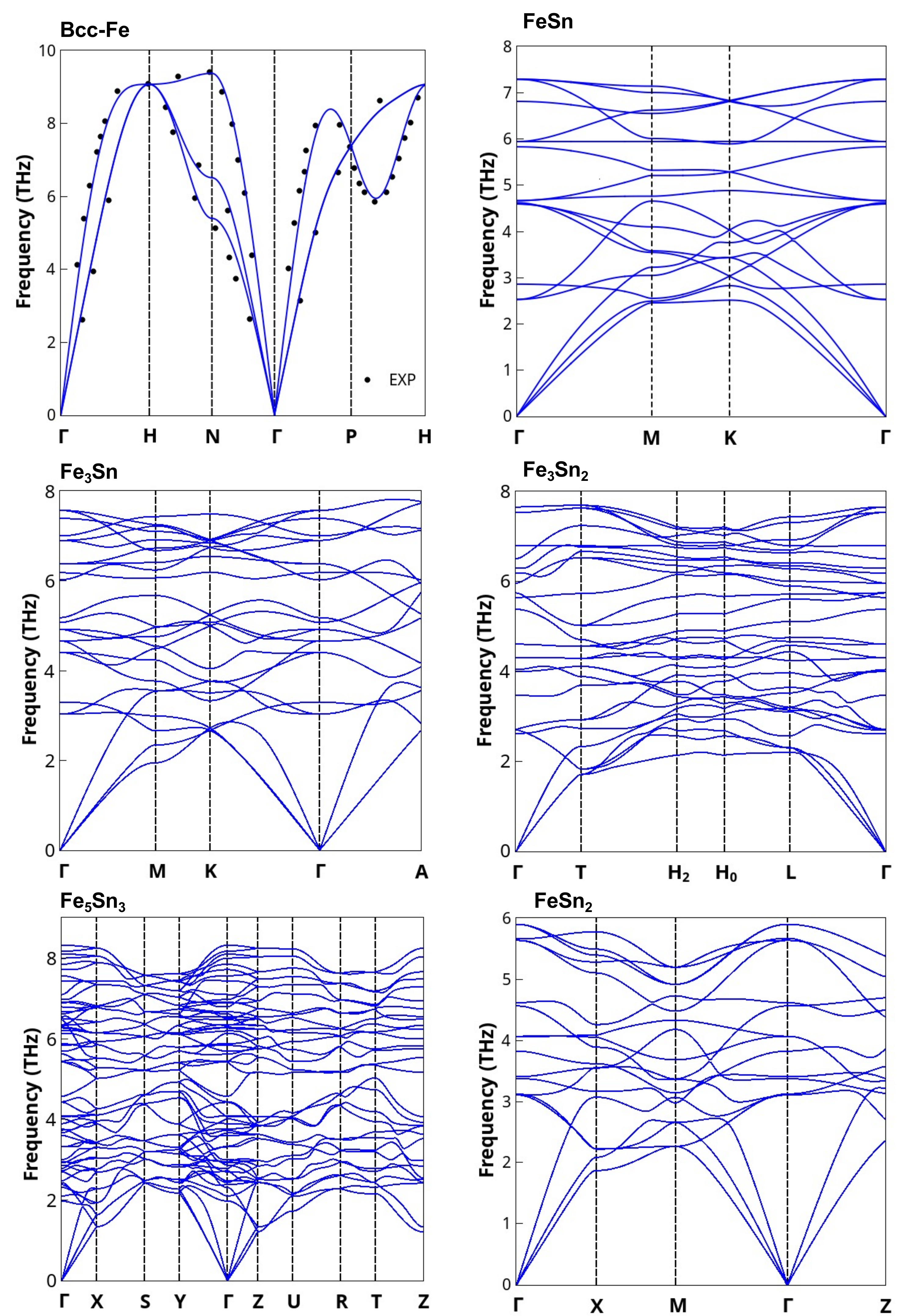}
    \caption{Phonon dispersions of the pure elements and intermetallic phases in the Fe-Sn system.}
    \label{phonon}
    \end{figure}

Most end-members in the sublattice models are not stable and their thermodynamic data are impossible to be determined by experiments. First-principles are hence performed to estimate the Gibbs energies of the compounds and end-members at finite temperatures.
In order to benchmark the current DFT calculations, the calculated crystallographic information of phases in the binary Fe-Sn system are listed in Table~\ref{lattice}, in comparison with the available experimental data. 
The calculated lattice parameters of the solid phases at 0 K are in good agreement with the experimental results at room temperature
As one can see, the differences between the theoretical and experimental lattice constants are within 0.5 $\%$ for all the phases. Note that, in our earlier study~\cite{fayyazi2017bulk,fayyazi2019experimental}, we showed, that the crystal structure of “Fe$_5$Sn$_3$” synthesized by the equilibrated alloy method, is not of the typically assumed hexagonal Laves structure (as shown in Table~\ref{lattice}). We rather observed superstructure reflections in the powder XRD spectra that could not be explained by the hexagonal structure and we assigned to a modulated orthorhombic unit cell with lattice parameters of a = 4.221 \AA, b = 7.322 \AA, c = 5.252 \AA. More details and explanations can be found in Ref.~\cite{fayyazi2017bulk,fayyazi2019experimental}. Hence, we used this structure to do phonon calculations. Furthermore, the calculated phonon bands of such phases are shown in Fig.~\ref{phonon}. 
To prove the validity of the calculations, as shown in Fig.~\ref{phonon}, the phonon dispersion of BCC-Fe is compared with the experimental data~\cite{brockhouse1967lattice}, presenting good agreement. 
Therefore, it is expected that the thermodynamical properties of the Fe-Sn intermetallic phases can also be accurately obtained based on DFT calculations.
As shown in Fig.~\ref{phonon}, no imaginary phonon modes exist for all the compounds, indicating that all the intermetallics are dynamically stable. And the quasi-harmonic approximation (QHA) can be used to calculate the thermodynamic properties.
    
\begin{figure}
    \centering   
    \includegraphics[width=12 cm]{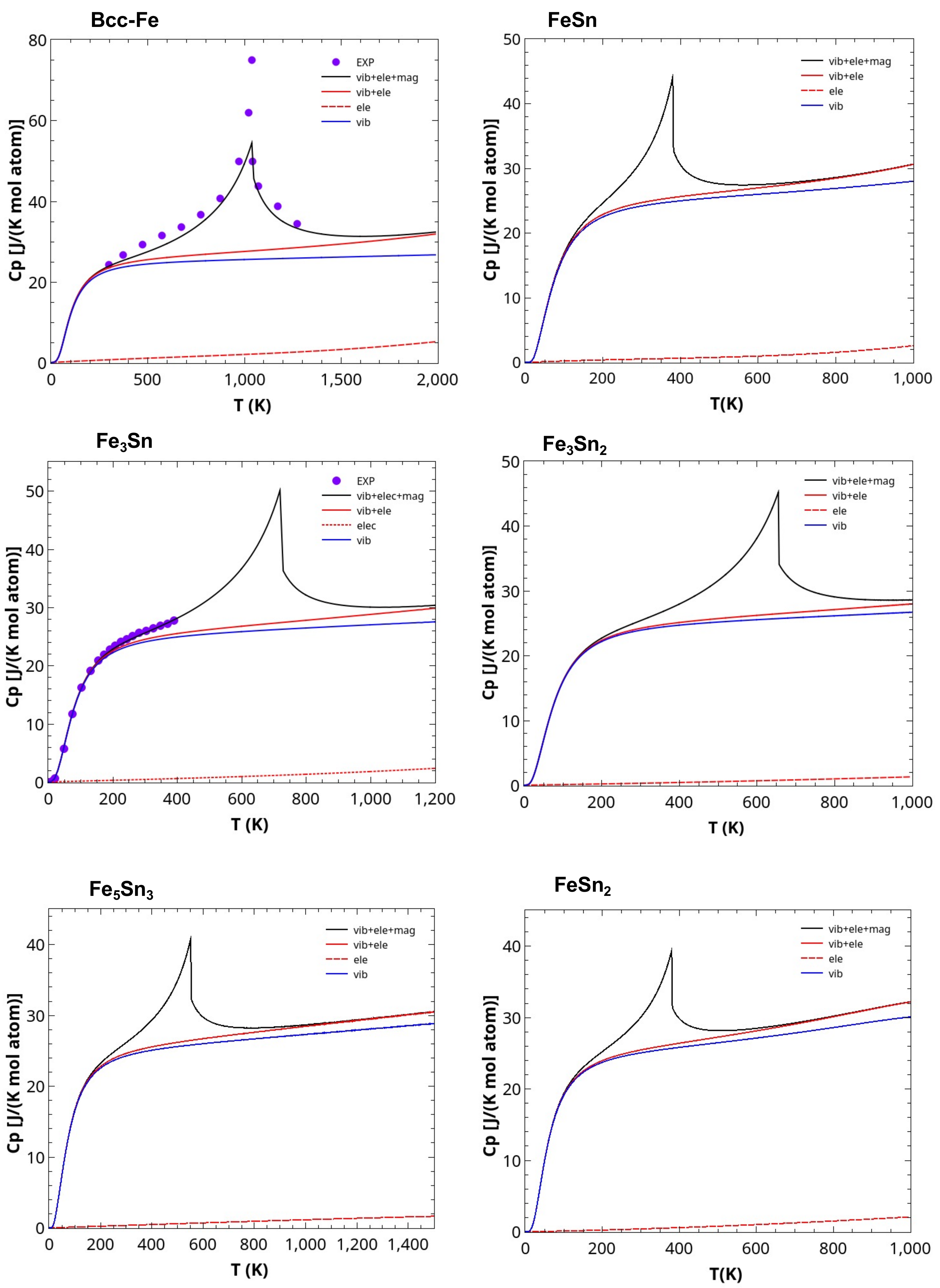}   
    \caption{Heat capacity of pure Fe and Sn from DFT calculations in comparison with the experiment data \cite{wallace1960specific,jennings1960lattice,berg1961high,novikov1974specific}, previous work \cite{zacherl2010first} and SGTE \cite{dinsdale1991sgte}. Those for all the intermetallics are also shown, experimental data of Fe$_3$Sn is obtained from our previous studies~\cite{fayyazi2017bulk,fayyazi2019experimental}.}
    \label{cp}
    \end{figure}

The thermodynamic properties at finite temperatures are evaluated based on the Gibbs free energies specified in Eq.~\ref{eq:GPT}. And from the thermodynamical point of view, we can derive the Gibbs free energies from the heat capacity.
To obtain the accurate heat capacity of the intermetallics, we firstly compare the calculated heat capacity of the BCC$\_$Fe with the available experimental data~\cite{wallace1960specific}, as shown in Fig.~\ref{cp}.
Among that, the magnetic contribution to the heat capacity is analyzed following the theory of Hillert and Jarl.\cite{hillert1978model}:
    \begin{equation}
    C_{p_{mag}} = Rln(\beta^{\varphi }+1)c(\tau ).
    \end{equation}
    
Fig.~\ref{cp} shows isobaric heat capacity obtained from our DFT calculations.
It can be found that the lattice vibrations to the heat capacity plays a dominant role. 
Interestingly, the correction made by adding
electronic and magnetic heat capacities shifted the result toward bigger values and after that calculations show an excellent agreement with the experimental data~\cite{wallace1960specific}.
More interestingly, the magnetic contribution to the heat capacity presents at the magnetic phase transition of BCC-Fe. 
These results prove the accuracy of the current methods and justify the following calculations for intermetallics.
Using the same strategy, we calculate heat capacities of Fe$_{5}$Sn$_{3}$, Fe$_{3}$Sn$_{2}$, Fe$_{3}$Sn, FeSn$_{2}$, and FeSn at finite temperatures, as shown in Fig.~\ref{cp}, with the magnetic heat capacity evaluated using Inden model~\cite{hillert1978model}.
The heat capacity of Fe$_{3}$Sn shows a good consistency between our calculations and experiments at low temperature, which also confirms the accuracy of current theoretical results. 
We note that such good agreements are supported by considering the magnetic contributions in the magnetic system.  

\begin{figure}[h!]
    \centering   
    \includegraphics[width=12CM]{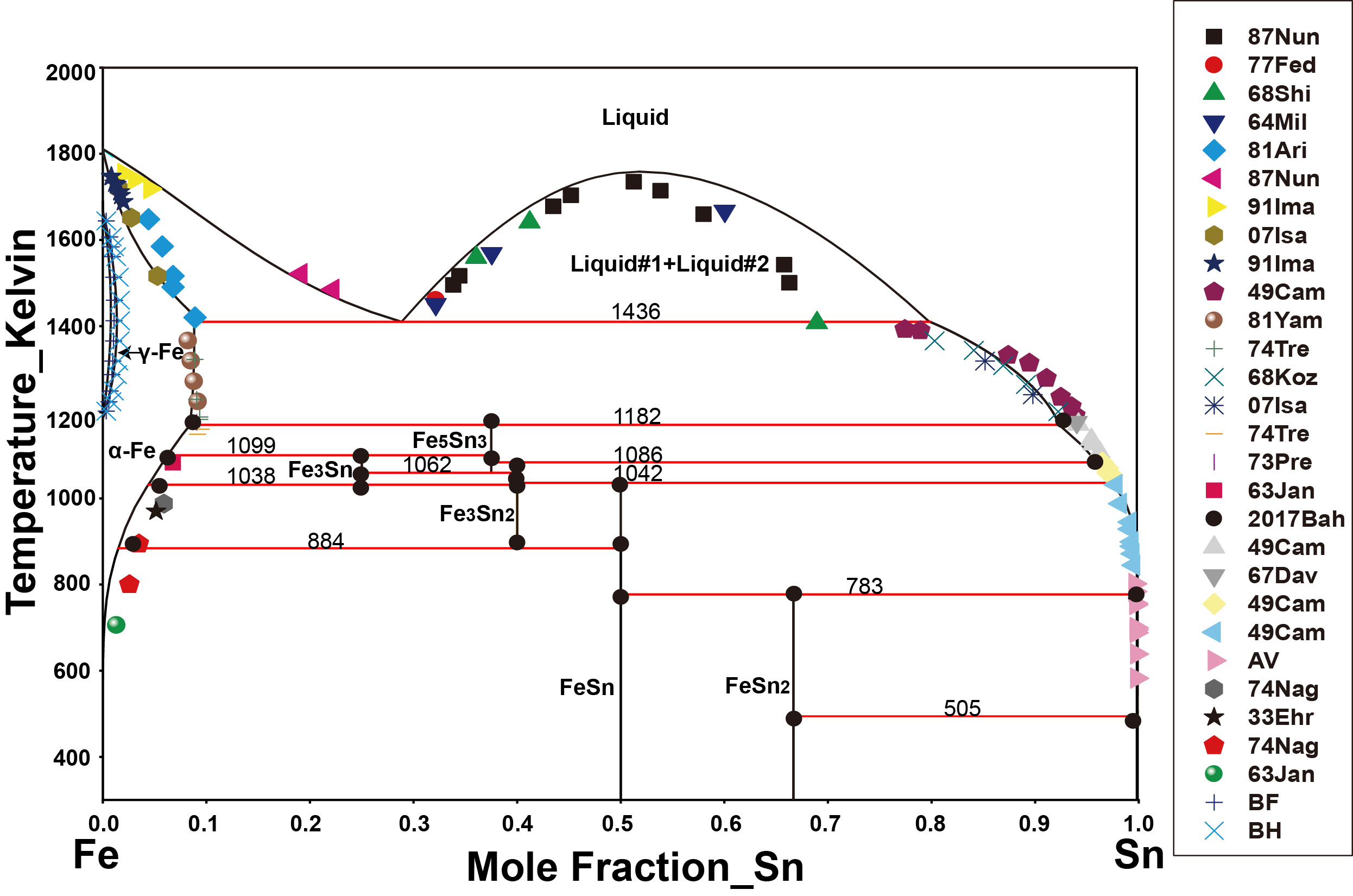}
    \caption{The optimized Fe-Sn phase diagram based on our thermodynamic modelling, in comparison with the experiment data~\cite{edwards1931,mills1964liquid,kubaschewski2013iron,isaac1907alloys,fedorenko1977vapor,treheux1974,singh1986,nunoue1987mass,mills1964liquid,arita1981measurements,massalski1991binary,yamamoto1981inter,jannin1963magnetism,fayyazi2017bulk}.}  
    \label{phase}
\end{figure}

After getting the thermodynamical properties of intermetallics, we used CALPHAD method~\cite{sundman2007computational} to evaluate the thermodynamic model parameters of the Fe-Sn system, and the phase diagram and thermodynamic properties are calculated by Thermo-Calc~\cite{sundman1985thermo}.Table S1 (see Supplementary) lists the modelled thermodynamic parameters of the Fe-Sn system.
The calculated Fe-Sn phase diagram is presented in Fig.~\ref{phase} along with the experimental data~\cite{edwards1931,mills1964liquid,kubaschewski2013iron,isaac1907alloys,fedorenko1977vapor,treheux1974,singh1986,nunoue1987mass,mills1964liquid,arita1981measurements,massalski1991binary,yamamoto1981inter,jannin1963magnetism,fayyazi2017bulk}. 
The comparison of the calculated temperatures and compositions of invariant reactions with experimental data~\cite{edwards1931,mills1964liquid,kubaschewski2013iron,isaac1907alloys,fedorenko1977vapor,treheux1974,singh1986} as well as results from previous thermodynamic assessments~\cite{huang2010sn,kumar1996} are listed in Table~\ref{reactions}.

\begin{table}[ht]
\caption{Summary of the invariant reactions in the Fe-Sn system.}
\tiny
\centering
\label{reactions}       
\begin{tabular}{lllllll}
\hline\noalign{\smallskip}
Invariant Reaction  &Reaction type& \multicolumn{3}{c}{Composition at \% Sn} & Temperature (K) & Refs.  \\
\noalign{\smallskip}\hline\noalign{\smallskip}
Liquid\#1 $\to$BCC\_A2+Liquid\#2  & Eutectic & 0.312& 0.083&0.811& 1395.9 & \cite{huang2010sn}\\
                                            &                &        &         &        &1381&\cite{edwards1931} \\
                                            &                &        &         &        &1404&\cite{mills1964liquid} \\
                                            &                &        &         &        &1403&\cite{kubaschewski2013iron} \\
                                            &                &        &         &        &1403&\cite{massalski1991binary} \\
                                            &                &        &         &        &1407&\cite{kumar1996} \\
                                            &                &        &         &        &1413&\cite{isaac1907alloys} \\
                                            & Eutectic               & 0.297       & 0.095        &  0.796      & 1436       &This work \\                                           

BCC\_A2 + Liquid $\to$Fe$_5$Sn$_3$&Peritectic&0.081 &0.929  &0.375 & 1174.1& \cite{huang2010sn} \\
                                            &                &        &         &        &1166&\cite{isaac1907alloys} \\
                                            &                &        &         &        &1168&\cite{kumar1996} \\
                                            &                &        &         &        &1183&\cite{treheux1974} \\
                                            &                &        &         &        &1183&\cite{fedorenko1977vapor} \\
                                            &  Peritectic              &  0.088      &   0.948      & 0.375       & 1182       &This work \\                                          

BCC\_A2 + Fe$_5$Sn$_3$ $\to$Fe$_3$Sn&Peritectic&0.062 &0.375 &0.250 &1099 & This work \\

Fe$_5$Sn$_3$ + Liquid $\to$Fe$_3$Sn$_2$    & Peritectic &  0.375     &0.967         &0.400       & 1074.8&\cite{huang2010sn}  \\
                                            &                &        &         &        &1072&\cite{singh1986} \\
                                            &                &        &         &        &1079&\cite{treheux1974} \\
                                            &                &        &         &        &1079&\cite{fedorenko1977vapor} \\
                                            &                &        &         &        &1080&\cite{kumar1996} \\
                                            &  Peritectic              & 0.375       &    0.979     &  0.400      & 1086       &This work \\   

Fe$_5$Sn$_3$ $\to$ Fe$_3$Sn + Fe$_3$Sn$_2$   
                                            &Eutectoid                &  0.375      &       0.250  &    0.400    &  1062      &This work \\  

Fe$_3$Sn$_2$ + Liquid $\to$FeSn  & Peritectic&  0.400             &0.980 &  0.500                &1024.7 &\cite{huang2010sn}   \\
                                             &                &        &         &        &1013&\cite{kubaschewski2013iron} \\
                                            &                &        &         &        &1034&\cite{kumar1996} \\
                                            &                &        &         &        &1043&\cite{treheux1974} \\
                                            &                &        &         &        &1043&\cite{fedorenko1977vapor} \\
                                            & Peritectic               &  0.400      & 0.979        &     0.500   &  1042      &This work \\          
                                            
Fe$_3$Sn $\to$ BCC\_A2 +Fe$_3$Sn$_2$    & Eutectoid &               0.250   & 0.046        &  0.400      &  1038      &This work \\     

Fe$_3$Sn$_2$ $\to$ BCC\_A2 +FeSn  & Eutectoid& 0.400    &0.017 & 0.500     &874.9 &\cite{huang2010sn}   \\
                                             &                &        &         &        &870&\cite{singh1986} \\
                                            &                &        &         &        &873&\cite{kubaschewski2013iron} \\
                                            &                &        &         &        &880&\cite{treheux1974} \\
                                            &                &        &         &        &880&\cite{fedorenko1977vapor} \\
                                            &                &        &        &        &880 &\cite{kumar1996} \\
                                            & Eutectoid               &  0.400      &  0.015        & 0.500       & 884       &This work \\

FeSn + Liquid $\to$FeSn$_2$  & Peritectic& 0.500              &0.999&   0.666               &775.4 &\cite{huang2010sn}   \\
                                             &                &        &         &        &769&\cite{isaac1907alloys} \\
                                            &                &        &         &        &769&\cite{edwards1931} \\
                                            &                &        &         &        &769&\cite{kubaschewski2013iron} \\
                                            &                &        &         &        &786&\cite{fedorenko1977vapor} \\
                                            &                &        &         &        &786 &\cite{treheux1974} \\
                                            &                &        &         &        &786 &\cite{kumar1996} \\
                                            & Peritectic               & 0.500       &0.999         &     0.666   &  783      &This work \\                                

Liquid $\to$ FeSn$_2$ +BCT\_A5  & Eutectic&0.999 &0.666 & 1.000                 &504.9&\cite{huang2010sn}   \\
                                             &                &        &         &        &501&\cite{isaac1907alloys} \\
                                            &                &        &         &        &505&\cite{fedorenko1977vapor} \\
                                            &                &        &         &        &505 &\cite{treheux1974} \\
                                            &                &        &         &        &505 &\cite{kumar1996} \\
                                            &  Eutectic              & 0.999       &  0.666       &        1.000& 505       &This work \\         
      
\noalign{\smallskip}\hline
\end{tabular}
\end{table}

Using the reactive crucible melting (RCM) approach, it is found that 3 metastable intermetallic compounds, i.e., Fe$_3$Sn, Fe$_5$Sn$_3$, and Fe$_3$Sn$_2$, can be stabilized between 873 K and 1173 K.
Furthermore, we are convinced that the phase diagram reported in the literature is inaccurate in the temperature interval 1023-1038 K and Fe$_3$Sn can exist at 1023 K.
Thus, the meta-stable phase Fe$_3$Sn is introduced by considering the current accurate experimental results.
Obviously, good agreement between the optimized and experimental Fig.S1-3 (see Supplementary) shows the calculated thermodynamic properties of the compounds in current CALPHAD modeling, first-principles calculations, previous CALPHAD modeling and the experimental data.
The calculated thermodynamic properties in this work are consistent with experimental data.

\begin{table}[ht]
\caption{Lattice parameters of intermetallics from first-principles calculations compared with experimental values.}
\label{lattice}       
\scriptsize
\centering
\begin{tabular}{lllccll}
\hline\noalign{\smallskip}
\multirow{2}{*}{Phases}  & \multirow{2}{*}{Space group} &\multirow{2}{*}{Magnetism} & \multicolumn{2}{c}{Lattice parameters ($\AA$)} & \multirow{2}{*}{k-point mesh}& \multirow{2}{*}{Refs.}  \\ 
              &                        &                              & \multicolumn{1}{c}{a}&\multicolumn{1}{c}{c}& & \\
\noalign{\smallskip}\hline\noalign{\smallskip}

 Fe$_3$Sn             & P6$_3$/mmc & FM &5.457&4.362&&\cite{buschow1983magneto}\\
                      &            &    &5.461&4.347&&\cite{cannon1984effect}\\
                      &            &    &5.421&4.434&&\cite{dobervsek1984metallographie}\\ 
                      &            &    &5.440&4.372&&\cite{kanematsu1986stability}\\ 
                      &            &    &5.464&4.352&&\cite{giefers2006high}\\  
                      &            &    &5.475&4.307&10$\times$10$\times$12&This work\\                          
 
Fe$_5$Sn$_3$& P6$_3$/mmc&FM&4.223 &5.253&&\cite{yamamoto1966mossbauer}\\  

Fe$_3$Sn$_2$& R-3 m&FM&5.344&19.845 &&\cite{malaman1976structure}\\ 
                 &               &        &5.340&19.797&&\cite{giefers2006high}\\ 
                 &               &        &5.315&19.703&&\cite{fenner2009non}\\
                &               &         &5.328&19.804&10$\times$10$\times$3&This work\\          
                
FeSn& P6/mmm&AFM&5.307&4.445&&\cite{buschow1983magneto}\\
                 &               &        &5.297&4.481&&\cite{van198657fe}\\
                 &               &        &5.288&4.420&&\cite{pavlyuk1997interaction}\\
                 &               &        &5.300&4.450&&\cite{waerenborgh2005crystal}\\ 
                 &               &        &5.298&4.448&&\cite{giefers2006high}\\
                 &               &        &5.297&4.449&&\cite{mikhaylushkin2008high}\\
                &               &         &5.299&4.449&10$\times$10$\times$10&This work\\       
                
FeSn$_2$&I4/mcm&AFM&6.502&5.315 &&\cite{ba1962neutron}\\ 
                &               &       &6.539&5.325&&\cite{havinga1972compounds}\\ 
                &               &       &6.539&5.325&&\cite{le1985mossbauer}\\
                &               &       &6.542&5.326&&\cite{van198657fe}\\
                &               &       &6.542&5.386&&\cite{pavlyuk1997interaction}\\
                &               &       &6.536&5.323&&\cite{giefers2006high}\\ 
                &               &       &6.533&5.320&&\cite{armbruster2007crystal}\\
                &               &       &6.545&5.326&&\cite{armbruster2010chemical}\\ 
                &               &       &6.561&5.338&8$\times$8$\times$10&This work\\                                        
\noalign{\smallskip}\hline
\end{tabular}
\end{table}

\subsection{Topological transport properties}
\begin{figure*}
\includegraphics[width=1\textwidth]{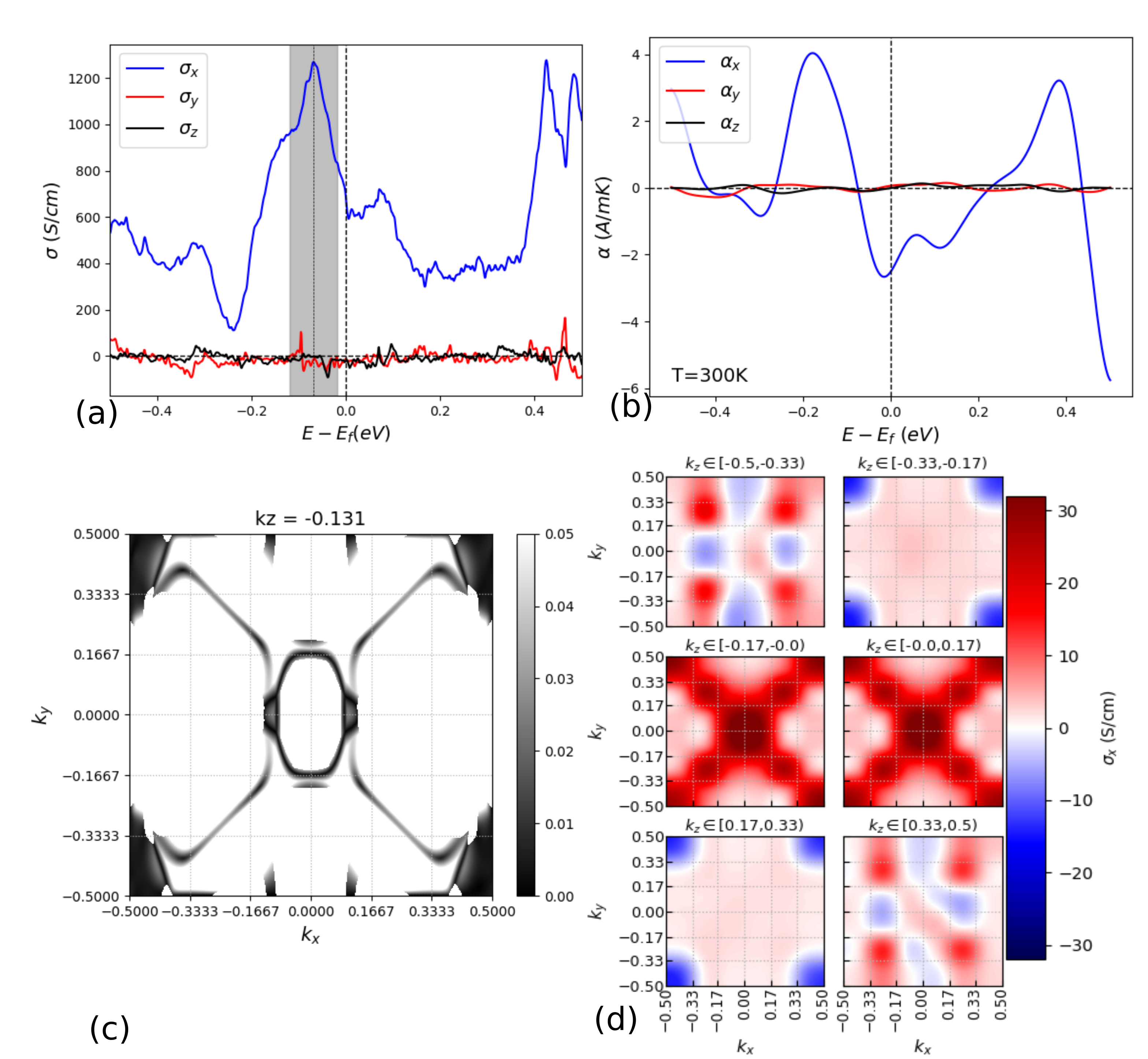}
\caption{(a) AHC components as a function of energy. (b) ANC components as a function of energy. (c) The band gap (eV) for a BZ-slice at $k_z=-0.131$ with its energies within the shaded energy range of part (a). (d) The distribution of the x-component ($\sigma_x=\sigma_{yz}$) of the full AHC at each part of the BZ.}
\label{fig1}
\end{figure*}

\begin{figure*}
\includegraphics[width=1\textwidth]{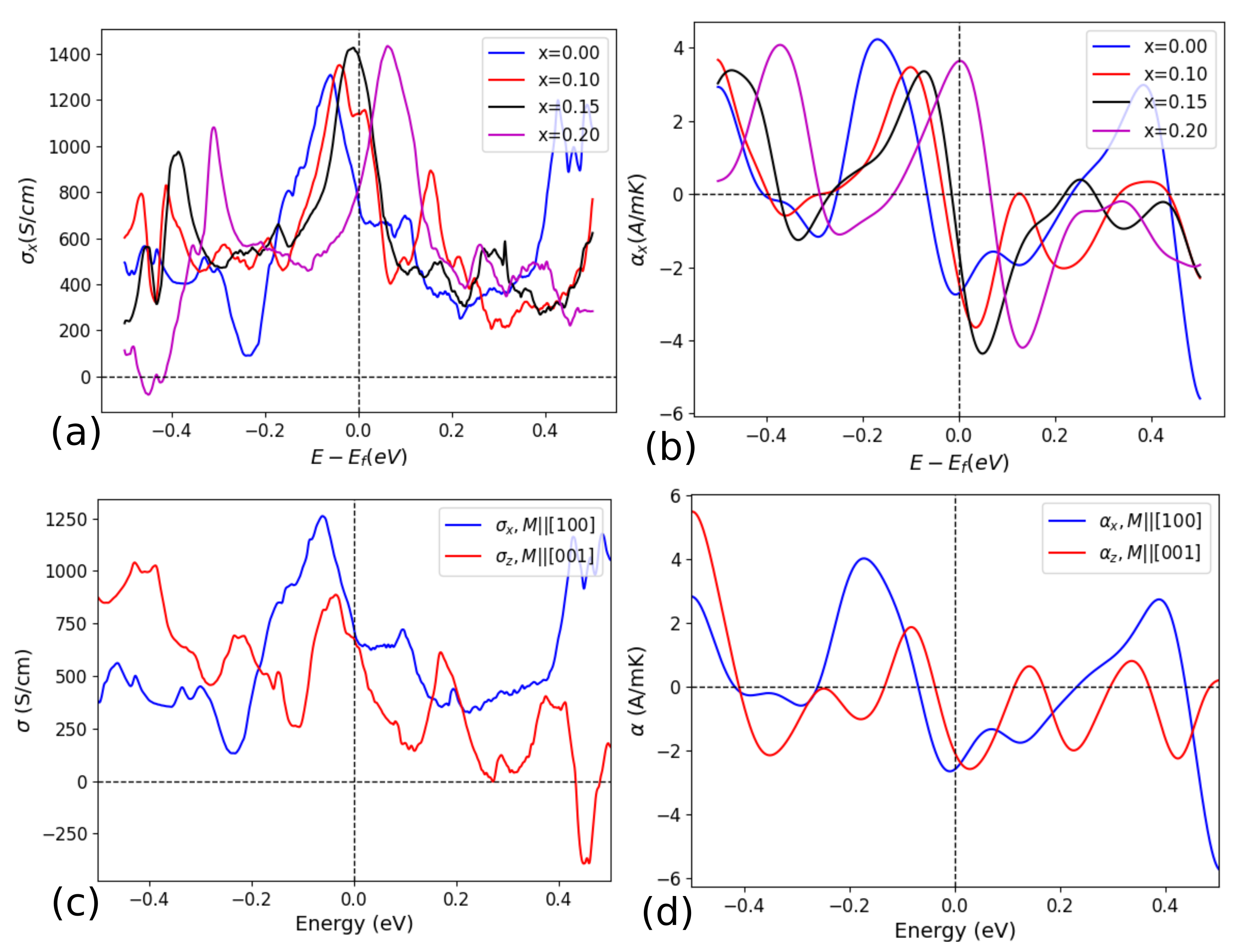}
\caption{(a) The x-component ($\sigma_x=\sigma_{yz}$) of the full AHC as a function of energy for different doping concentrations x of (Fe$_{1-x}$Mn$_{x}$)$_{3}$Sn. (b) The x-component ($\alpha_x=\alpha_{yz}$) of the full ANC as a function of energy for different doping concentrations x of (Fe$_{1-x}$Mn$_{x}$)$_{3}$Sn. (c) AHC as a function of energy for different magnetization directions of Fe$_3$Sn. (d) ANC as a function of energy for different magnetization directions of Fe$_3$Sn.}
\label{fig2}
\end{figure*}

Our calculations demonstrate that Fe$_3$Sn exhibits the largest AHC among the Fe-Sn family. The calculated x-component of AHC ($\sigma_x$) for the equilibrium lattice parameters, with the magnetization direction along [100]-axis, reaches 757 $S/cm$ at Fermi energy, as shown in Fig.~\ref{fig1}(a). This value is surprisingly large compared to the more than 3 times smaller value of 200 $S/cm$ for Fe$_3$Sn$_2$.~\cite{ye2018massive}. Additionally, Li et al. reported an experimentally measured value of 613 $S/cm$ value as well as a calculated of 507 $S/cm$ for Fe$_5$Sn$_3$~\cite{li2020large}, both being smaller than our value for Fe$_3$Sn. In addition, compared to other ferromagnetic kagome materials, it ranks among the largest reported, being larger than 380 $S/cm$ of LiMn$_6$Sn$_6$~\cite{chen2021large} and 223 $S/cm$ for GdMn$_6$Sn$_6$~\cite{asaba2020anomalous}, but lower than the largest reported value of 1130 $S/cm$ for Co$_3$Sn$_2$S$_2$~\cite{liu2018giant}. Correspondingly, the ANC of Fe$_3$Sn, evaluated at $T=300K$, is also the largest among the Fe-Sn family. Specifically, it reaches -2.71 $A/mK$ (see Fig.~\ref{fig2}(b)) being more than 2 times larger than the reported value of 1 $A/mK$ for Fe$_3$Sn$_2$~\cite{zhang2021topological}. Compared to the other kagome materials, Fe$_3$Sn exhibits a reasonably large ANC, being larger than 1.29 $A/mK$ and 0.20 $A/mK$ reported for ZrMn$_6$Sn$_6$ and MgMn$_6$Sn$_6$~\cite{samathrakis2021enhanced}, but smaller than the largest reported value of 10 $A/mK$ for Co$_3$Sn$_2$S$_2$~\cite{yang2020giant}. Compounds with large AHC and ANC values are promising candidates for transverse thermoelectric devices using geometries~\cite{zhou2021seebeck,uchida2021transverse,sakai2020iron}.

Symmetry plays a crucial role in determining the shape of the AHC and ANC tensors. AHC and ANC strongly depend on the Berry curvature which behaves as a pseudovector under the application of any symmetry operation~\cite{seemann2015symmetry,Suzuki:2017} and transforms according to the formula
\begin{equation}
s\mathbf{\Omega\left(r\right)}=\pm det \left( \mathbf{D}\left(R\right) \right) \mathbf{D}\left(R\right) \mathbf{\Omega}\left(s^{-1}\mathbf{r}\right), \notag
\end{equation}
where $\mathbf{\Omega\left(r\right)}$ denotes the pseudovector Berry curvature, $\mathbf{D}\left(R\right)$ the three-dimensional representation of a symmetry operation without the translation part and s an arbitrary symmetry operation. That is, the symmetry operations of the magnetic point group will govern the shape of the tensors. Particularly, the ferromagnetic Fe$_3$Sn belongs to the magnetic space groups $Cmc'm'$ (BNS: 63.463), $Cm'cm'$ (BNS: 63.464) and $P6_3/mm'c'$ (BNS: 194.270) for the magnetic moments of Fe atoms pointing along the [100]-, [010]- and [001]-axis, respectively. Hence the presence of the $2_x$, $2_y$ and $2_z$ rotation axes for each of magnetic space groups transform the Berry curvature according to:
\begin{equation}
\begin{split}
& \text{For } 2_x \text{ with } M||[100] \\
& \Omega_{x}\left(k_x,-k_y,-k_z\right) = \Omega_{x}\left(k_x,k_y,k_z\right) \\
& \Omega_{y}\left(k_x,-k_y,-k_z\right) = -\Omega_{y}\left(k_x,k_y,k_z\right) \\
& \Omega_{z}\left(k_x,-k_y,-k_z\right) = -\Omega_{z}\left(k_x,k_y,k_z\right). \\
& \\
& \text{For } 2_y \text{ with } M||[010] \\
& \Omega_{x}\left(-k_x,k_y,-k_z\right) = -\Omega_{x}\left(k_x,k_y,k_z\right) \\
& \Omega_{y}\left(-k_x,k_y,-k_z\right) = \Omega_{y}\left(k_x,k_y,k_z\right) \\
& \Omega_{z}\left(-k_x,k_y,-k_z\right) = -\Omega_{z}\left(k_x,k_y,k_z\right). \\
& \\
& \text{For } 2_z \text{ with } M||[001] \\
& \Omega_{x}\left(-k_x,-k_y,k_z\right) = -\Omega_{x}\left(k_x,k_y,k_z\right) \\
& \Omega_{y}\left(-k_x,-k_y,k_z\right) = -\Omega_{y}\left(k_x,k_y,k_z\right) \\
& \Omega_{z}\left(-k_x,-k_y,k_z\right) = \Omega_{z}\left(k_x,k_y,k_z\right). \\
\end{split}
\end{equation}
The summation over the whole Brillouin zone forces $\sigma_y$ and $\sigma_z$ for the magnetization direction along the [100]-axis to vanish, and equivalently $\sigma_x$ and $\sigma_z$ and $\sigma_x$ and $\sigma_y$ for the magnetization along [010] and [001], respectively. However, there is no such condition for $\sigma_x$, $\sigma_y$ and $\sigma_z$ for magnetization direction along [100], [010] and [001] axes respectively, and therefore they are allowed to have finite values.

AHC and ANC are proportional to the sum of the Berry curvature of the occupied bands, evaluated in the whole Brilouin zone (BZ), as defined in Eq~\ref{bc-ahc}. Since the Berry curvature depends on the energy difference between two adjacent bands, therefore it is expected that Weyl nodes as well as nodal lines, located close to the reference energy, contribute significantly to the total value, as shown in Ref.~\cite{nagaosa2010anomalous,xu2020high} and confirmed for MnZn~\cite{samathrakis2021enhanced} and Mn$_3$PdN~\cite{singh2021multifunctional}, respectively. Explicit band structure search reveals the presence of numerous Weyl nodes and nodal lines within the shaded energy range [-0.118,-0.018]eV of Fig~\ref{fig1}(a) that are expected to contribute to the total AHC value.
In order to identify the origin of the AHC contribution, we split the BZ into 216 cubes, within which the AHC is evaluated (see Fig.~\ref{fig1}(d)). Since the major contribution originates from the diagonals, located within $k_z\in\left(-0.166,0.000 \right)$ (and $k_z\in\left(0.000,0.166 \right)$), as illustrated in Fig.~\ref{fig1}(d), it is fruitful to investigate the band gap within this $k_z$ range. Taking as an example the $k_z=-0.131$ plane, we plot the difference of the two involved bands as a black and white plot where the black areas correspond to small gap regions whereas white areas to large gap regions (see Fig.~\ref{fig1}(c)). The shape of the gap plot is in complete agreement with the distribution of the AHC within the specified area, demonstrating that small gap regions similar to those within the square $k_x,k_y\in\left(0.333,0.500\right)$ of Fig.~\ref{fig1}(c), contribute dominantly to the total AHC value.

Interesting topological transport properties can arise away from the charge neutral point. One important observation is that the AHC curve of Fe$_3$Sn exhibits a sharp peak of 1308 $S/cm$ located at 60 $meV$ below the Fermi level, as shown in Fig.~\ref{fig1}(a). Therefore, it is reasonable to wonder whether tuning the Fermi level to match the position of the peak is doable by means of doping. In order to investigate this possibility, we consider (Fe$_{1-x}$Mn$_{x}$)$_{3}$Sn for various values of $x$, with $x\in\left[0,0.2\right]$, indicating the percentage of Mn doping to the system. By using virtual crystal approximation (VCA) calculations, we compute the AHC curve for different $x$, as illustrated in Fig.~\ref{fig2}(a). It is noted that the position of the peak approaches the Fermi level while the Mn dopand concentration is increased and it hits the Fermi energy at approximately $x=0.15$ (black curve). The existence of the AHC peak and its location affects the calculated ANC. While the energy of the peak is lower than the Fermi energy ($x<0.15$), the ANC is gradually decreased from -2.71 $A/mK$ for $x=0$ to -1.58$A/mK$ for $x=0.15$. Once the energy of the peak gets larger than the Fermi energy ($x>0.15$), ANC changes sign and jumps to 3.63 $A/mK$ for $x=0.2$. Fig.~\ref{fig2}(b) shows the calculated ANC curves for various x, demonstrating that Fe$_3$Sn offers an interesting playground of controlling the ANC by doping even with a sign change.

Tuning the magnetization direction allows easier ANC modifications. In an attempt to tune the AHC and ANC of Fe$_3$Sn, we considered the magnetization direction as a perturbation by aligning it along [100], [010] and [001] axes. Our results show no impact of the magnetization direction to the AHC and ANC values along [100] and [010] axis, where the values remain practically unchanged at 757 $S/cm$ and -2.58 $A/mK$ due to the underlying hexagonal symmetry. On the other hand, a small change is noticed for direction along [001], where the AHC (ANC) is tuned to 676 $S/cm$ (-2.06 $A/mK$), see Fig.~\ref{fig2}(c) and (d). Despite the minor changes in the AHC and ANC values at Fermi energy, a larger impact of the altering of the magnetization direction is observed away from the charge neutral point. Specifically, the peak of 1308 $S/cm$ at 60 $meV$ below the Fermi energy is moved closer to the Fermi energy, at 35 $meV$ below the Fermi energy, and further reduces its maximum value to 886 $S/cm$ when the magnetization direction is along the [001]-axis. The outcome of this change is more obvious in the ANC, where the zero value of the [001] direction is located closer to the Fermi energy, being useful for future applications. 

\section{\label{CONC}Conclusion}

Based on DFT calculations, the thermodynamical properties of the Fe-Sn system and topological transport properties of Fe$_3$Sn are studied. Thermodynamic modeling of the Fe-Sn phase diagram has been re-established. The problems concerning invariant reactions of intermetallics are remedied under our newly measured temperature ranges. First-principles phonon calculations with the QHA approach were performed to calculate the thermodynamic properties at finite temperatures. Thermodynamic properties, phonon dispersions of pure elements and intermetallics were predicted to make up the shortage of experimental data. The heat capacity of Fe$_3$Sn is measured for the first time, the values match well with the predicted results, confirming the accuracy of our DFT calculations. A set of self-consistent thermodynamic parameters are obtained by the CALPHAD approach. Further, we evaluated the AHC and ANC of Fe$_3$Sn with magnetization direction and doping being perturbations. The calculated AHC of 757 $S/cm$ is the largest among all reported members of the Fe-Sn family. It is noted that the nodal lines combined with the extended small gap areas constitute the main contribution to the total AHC and they can further be tuned by doping Mn at the Fe sites, allowing the manipulation of the AHC and ANC values and offering good candidate materials for promising transverse thermoelectric devices.

\begin{center}
\small \textbf{ACKNOWLEDGMENTS}
\end{center}

This work was financially supported by the Deutsche Forschungsgemeinschaft (DFG) via the priority programme SPP 1666 and the calculations were conducted on the Lichtenberg high performance computer of the TU Darmstadt.
\clearpage

\end{document}